\documentstyle[preprint,eqsecnum,aps,floats]{revtex}
\begin{document}
%
\begin{titlepage}
\preprint{NSF-ITP-96-151, UCSBTH-96-30}
\title{Sources of Predictability\thanks{To appear in the Proceedings
of the conference on {\sl Fundamental Sources of Unpredictability}
held at the Santa Fe Institute, March 28 to 30, 1996 to be published
by {\sl Complexity}}}
\author{James B.~Hartle\thanks{hartle@itp.ucsb.edu}}
\address{Institute for Theoretical Physics,\\
University of California, \\
Santa Barbara, CA, 93106-4030}

\date{\today}
\maketitle

\tighten
\begin{abstract}

Sources of predictability in the basic laws of physics are
described in the most general theoretical context --- the quantum
theory of the universe as a whole. 

\end{abstract}

\end{titlepage}

\tighten
\section{Introduction}

The title of our conference, {\sl Fundamental Sources of Unpredictability},
invites the question: ``What are the obstacles and limitations to the
predictive power of the laws of physics?''  But in a quantum mechanical
world characterized by indeterminacy and distributed probabilities, in a
world where even classically describable phenomena can be intractable
or impossible to compute, and in a finite universe with limited
opportunities for observation and induction, we might more straightforwardly
ask the question: 
``What are the fundamental
origins of predictability?''  More specifically, 
what features of the basic laws of physics lead to predictable regularities
in the universe?
This essay is devoted to a few general remarks
about this question from the perspective of quantum cosmology --- the
quantum theory of the universe as a whole.

It seems inescapable  that there are limits to the predictive power of the
laws of physics. The world appears to be complex. But the basic laws
that govern the regularities of the world must be simple 
to be discoverable, comprehensible, and effectively applicable.  It is a
logical possibility that every feature of our experience --- the shape
of each galaxy, the number of planets circumnavigating each star, the
character of each biological species, the course of human history, and
the result of every experiment --- is the output of some short computer
program with no input. However, there is no evidence that our universe
is so regular. Even the most deterministic classical theories did not
claim this.  They claimed only to predict future evolution when {\it given}
initial conditions.  Not everything that can be observed can be
predicted --- only the regularities in those observations are the province of
science. The regularities of the universe are limited  and therefore
there are limits to the predictive power of physics. This kind of
limit is therefore not a failure 
of the scientific enterprise.  Limits are inherent to that enterprise,
and their demarcation is an important scientific question.

To ask about the {\it fundamental} sources of
predictability is to ask for the regularities implied by the basic laws
that apply universally to all physical systems --- without exception,
qualification, or approximation --- in the most general theoretical context.
In contemporary physics,  that is the subject of quantum cosmology --- the
quantum theory of the initial condition of the universe and its
subsequent history.  In present thinking this theory is based on two
laws:
\begin{itemize}
\item The basic theory of dynamics ({\it e.g.}~heterotic superstring theory),
\item The theory of the initial condition of the universe ({\it e.g.}~the
``no-boundary'' wave function of the universe).
\end{itemize}

There are no interesting predictions that do not depend on these two
laws, even if only very weakly.  
In this essay, we shall describe some specific features of these two 
basic laws which give rise to predictable general regularities of
the universe. We cannot pretend to discuss the
origin of all the regularities considered by science in this brief
compass, but we 
shall describe some of the more general ones. We begin in Section II
with the general regularities exhibited
by any quantum system, and then proceed to the regularities that 
depend on the particular properties of the laws of dynamics and initial
condition that hold in our universe. Specifically, we consider the 
origin of a realm of approximate classical determinism in Section III
and the origin of individual, approximately isolated subsystems in
Section IV. Mere existence of such regularities, however, is not 
enough for science. They must be accessible to inference and 
amenable to computation. We mention a few properties of the basic
laws that allow for these qualities in Sections V and VI. 

\section{A Quantum Universe}

It is an inescapable inference from the physics of the last seventy years 
that we live in a quantum mechanical universe and that the basic laws of
dynamics and initial condition are consistent with the general
framework of quantum theory. Let us accept the inference and ask for its
implications for predictability.

To keep the discussion manageable  let us neglect gross
quantum fluctuations in the geometry of spacetime such as would be
expected to occur in the first $10^{-43}$ sec.~after the big bang. 
The quantum mechanics of a universe of matter fields moving in
a fixed spacetime geometry exhibits the determinism of the Schr\"odinger
equation:
\begin{equation}
i\hbar\ \frac{\partial\Psi}{\partial t} = H\Psi \ . 
\label{twoone}
\end{equation}
The basic theory of dynamics supplies the Hamiltonian $H$, the theory 
of the initial condition supplies the initial 
quantum state $\Psi_0$.
The only regularities 
exhibited with certainty by all quantum systems, independent of the
specific natures of $H$ and $\Psi_0$, are 
paraphrases of the prediction that at any time $t$ the universe is in
the state $\Psi(t)$ that evolved deterministically
from the initial state $\Psi_0$ by the
Schr\"odinger equation. 

But it is not merely these  certain
and general regularities of the Schr\"odinger equation which are 
of interest. 
We are interested in the much broader class of regularities 
which are not certain but occur with high probability as a consequence
of the {\it particular} properties of the laws of dynamics and initial
condition. The approximate regularities of classical physics are
the most prominent example. 

Most generally quantum mechanics predicts the probabilities of
particular {\it histories} of the universe in sets of alternative
possible histories. Examples of sets of alternative histories are the
alternative histories of the abundances of the elements 
over time,
the alternative histories of the motion of the earth around the sun, the
alternative histories of outcomes of a laboratory experiment testing
quantum mechanics, the alternative evolutionary tracks of biological
species on earth, and the alternative histories of the S\&P stock
average.  Probabilities for the members of each of these sets of
alternative histories can in principle be calculated from the initial
state of the universe and the basic dynamical law. 

Probabilities of single events or single histories are not necessarily definite predictions. From a
theory that a coin is unbiased, we calculate the probability of the
outcome of a single toss as $1/2$, but we cannot be said to have made a
definite prediction of the outcome. 
The definite predictions of a theory of the initial state and dynamics
are the sets of histories in which only one member has a probability very close
to unity and the rest have probabilities close to zero. Among the examples of sets of alternative histories mentioned above, only the first ---
the alternative histories of the abundances of the elements --- has this
character.  Current theories of the initial state and the dynamics
predict a high probability of primordial abundances of approximately 75\%
hydrogen, 25\% ${\rm He}^4$, small but definite abundances of ${\rm He}^3$,  deuterium,
lithium, and negligible percentages of the other elements.
There are similarly definite predictions for some other large scale
features of the universe, 
but the probabilities for the other sets of alternatives mentioned above
are likely to be highly distributed and so are not definite 
predictions. 
We expect only a few definite predictions from the basic
laws of dynamics and initial condition alone, unaugmented by further
information. 
The largest and most useful general regularities
of this kind are those summarized by the deterministic laws of 
classical physics.  It is to these we now turn. 

\section{The Quasiclassical Realm}

Classical deterministic laws approximately govern the regularities in
time of a wide range of
phenomena over a broad span of time, place, and scale in the universe.
The domain of applicability of these classical laws 
is the quasiclassical realm of everyday
experience.  In a quantum mechanical universe these classical equations
can be but approximations to unitary
evolution by the Schr\"odinger equation and reduction of the state vector.
To what do we owe
the validity of this classical approximation over such an extensive realm of
phenomena, time, place, and scale?

The answer is certainly not merely the determinism of the Schr\"odinger
equation. The regularities in time that it summarizes generally relate
quantities that are nothing like those of classical physics. The simple
example of Ehrenfest's theorem for the quantum mechanics of a single particle
moving in one dimension suggests what is necessary for classical
behavior.  Ehrenfest's theorem relates the acceleration of the expected
position to the expected value of the force:
\begin{equation}
m\ \frac{d^2\langle x\rangle}{dt^2} = - \left\langle\frac{\partial
V(x)}{\partial x}\right\rangle \ .
\label{threeone}
\end{equation}
This is an exact consequence of the determinism of the Schr\"odinger
equation, but not a classical 
equation of motion because it is not a differential equation for $\langle
x\rangle$. It becomes a deterministic equation in the approximation
that the expected value of the force may be replaced by the force
evaluated at the expected position
\begin{equation}
m \frac{d^2\langle x\rangle}{dt^2} \approx - \frac{\partial
V(\langle x\rangle)}{\partial x} \ .
\label{threetwo}
\end{equation}
There are at least four requirements for the validity of this
approximation. First, the Hamiltonian must be of the form that can
give rise to an equation of motion. Second, the approximation is true {\it only for certain states},
typically narrow wave packets. Third, the right variables must be
followed, in this case position. Finally, there must be coarseness in how
they are followed, here average rather than exact position.

Only certain states and certain coarse-grained descriptions exhibit
patterns of classical correlations in this simple model. Similarly, 
but more generally, we cannot expect a quasiclassical realm in
a universe with a generic initial condition and a generic Hamiltonian. 
The approximate quasiclassical realm that extends over most of
cosmological space and time in this universe owes its existence to 
an appropriate
coarse-grained description in terms of the variables of classical
physics and the particular properties of the initial quantum state
and dynamical law.
Classical predictability is an emergent feature of the universe's
particular initial condition and dynamical law.

\section{ Regularities of Subsystems}

Most of science is not concerned with the regularities exhibited
universally by all physical systems but rather with the particular
regularities of classes of individual, localized, subsystems. Stars, atoms, oceans, biological
species and individual human behaviors are just a few of the myriad
examples. In interesting circumstances such subsystems approximately
preserve their identity over significant periods of time and can be treated as
approximately isolated. Their properties and evolution can thus be studied
individually rather than as a feature of the evolution of the whole
universe. There is no particular reason to believe that a universe
with a generic Hamiltonian and a generic initial state would exhibit
such approximately isolated subsystems. To what features of our particular
basic laws do we owe the existence
of such subsystems and what is the origin of their regularities?

The locality of the effective interactions between the elementary particles
that is a consequence of the basic dynamical law is certainly necessary
for the approximate isolation of localized subsystems. However, 
properties of the law of initial condition also play a role.
There were no individual subsystems in the early universe. The 
evidence of the observations is that the early universe was
nearly smooth, featureless, homogeneous,  isotropic, and far from
equilibrium. Today's subsystems
have evolved from the quantum fluctuations away from this smoothness
that are properties of the initial quantum state
primarily by the universal action of gravitational attraction.
The existence of individual galaxies, for example, is a consequence
of the gravitational collapse of primordial fluctuations in the
uniform density of matter.

There are several possible sources for the regularities among members
of certain classes of approximately isolated subsystems. The simplest
is a common historical origin as with individual members of a biological species.
But the laws of dynamics and initial condition can also be a source
of regularities. Einstein's theory of gravity ensures that one 
black hole is similar to any other no matter how different were 
the situations in which they were formed. 
The regularities among galaxies arise
from their origin in similar circumstances across the universe as
mandated by an initial condition of close to perfect homogeneity
and isotropy.

Thus the specific regularities of particular subsystems that are
the subject of much of science outside of physics arise in part 
from special properties of the basic laws of dynamics and the initial
condition of the universe.  
\section{Inference and Computation}

In the previous sections we have discussed how the fundamental laws of
dynamics and the initial condition permit prediction in this quantum
mechanical universe.  But these laws are not presented to us by
revelation. They are arrived at by a process of induction from
observation, theoretical development, and experimental test. Further,
exhibiting the predictions of these laws is not simply a matter of
displaying their form. Their consequences must be computed in particular
circumstances.  No treatment of the sources of predictability would be
complete without  discussion of those features of the two fundamental laws
that permit the processes of induction and application.  
Space does not permit an examination of these questions that is worthy of
their depth. We can only list a few features of the fundamental laws that
help us in our task.

The locality of the fundamental law of dynamics, and the separation of
phenomena by scale that it permits,  have both aided its discovery. Locality
enables us to extrapolate the regularities of familiar scales both much
larger and smaller domains.  Assuming locality, 
from gravitational phenomena on earth we are able to infer the rules governing the evolution
of the planets, galaxies, and eventually the universe itself. Assuming
locality, from the electromagnetic forces between
macroscopic bodies we are able to infer those between the constituents of 
atoms.  

Separation of phenomena by scale has permitted a kind of progression in
discovery of the fundamental dynamical law --- each stage being
characterized by a discoverable effective theory of limited validity.
Thus we moved from the motion of the planets in the solar system, to the
motion of electrons in an atom, to the motion of protons and neutrons in the
nucleus, to the motion of quarks in the proton and neutron, {\it etc., etc. } Thus 
classical physics was discovered before quantum mechanics, Newtonian
mechanics before special relativity, and Newtonian gravity before
general relativity, {\it etc}. The similarity of phenomena on different scales has
been advanced \cite{GMup} as the reason for the ``unreasonable effectiveness of
mathematics in the natural sciences.'' 

Locality of effective interactions and separation of phenomena by scale 
also reduce the computation of certain predictions to manageable
tasks. Locality has already been mentioned as a prerequisite for
approximately isolated subsystems. For these the application of the
dynamical law leads to computations which are more tractable than
those arising in its application to the universe as a whole. The separation of
phenomena by scale permits the computation of the evolution of a wide
range of phenomena using
simpler, approximate, effective forms of the dynamical law. 

More generally it is important for practical predictability that the
known laws of physics predict computable numbers extractable in
particular circumstances by the application of standard algorithms. 
This is not a trivial statement because
there are infinitely many more non-computable numbers than computable
ones.

The phenomenon of chaos is commonly mentioned as a source of
unpredictability in physics. But in an essay on sources of
predictability it is more appropriate to focus on the existence of
non-chaotic, integrable systems whose evolution can be tractably 
computed.  The
source of this integrability lies in the conservation laws of the
dynamical theory and these arise from its symmetries --- exact or
approximate. These same conservation laws are at the heart of the
effectiveness of statistical mechanics in 
separating out quantities that
approach equilibrium slowly enough to be governed by phenomenological
equations of motion. In these ways symmetries of the dynamical law can
be a source of predictability.

\section{Conclusion}

Predictable regularities in this specific quantum universe arise from
particular features of its 
fundamental laws of dynamics and the initial condition. As
observers of the universe we rely on the existence of a quasiclassical
realm, the locality of effective interactions, the
separation of phenomena by scale, and the existence and regularities
of approximately isolated subsystems --- all properties which depend
on the form of the basic laws of dynamics and initial condition. One 
could ask: ``Why do the fundamental laws exhibit
these features which are of such use to us?'' That, however, is the wrong
way of putting the question. Individually and collectively we are
complex adaptive systems within the universe that evolved to exploit the
emergent regularities that the universe presents. Ask rather how or
whether such systems would have evolved if the fundamental laws did not
display the features we have described. While much beyond our power to
answer at present, this question does show that our adaptation to the
regularities of our specific universe is another source of
predictability in physics.

\section{Further Reading}
A {\sl Scientific American} introductory article to quantum cosmology
is \cite{HalXX}. The author has expanded on many of the topics in this
brief essay in \cite{HarYY} where a more extensive list of references
may be found.

\acknowledgments

Preparation of this essay was supported in part by the US National
Science Foundation under grants PHY95-07065 and PHY94-07194.

\end{document}